\documentclass[a4paper,12pt]{article}
\usepackage[utf8]{inputenc}
\usepackage{cancel}
\usepackage{ulem}
\usepackage{amsfonts}
\usepackage{amssymb}
\usepackage{graphicx}
\usepackage{amsmath}
\usepackage{enumerate}
\usepackage{mathtools}
\usepackage{subfig}
\usepackage{color}
\usepackage{tikz}\usetikzlibrary{calc}
\usepackage{float}
\usepackage{here}
\usepackage{cite}
\usepackage{mathrsfs}
\usepackage{float,epsfig}
\usepackage{dcolumn}

\usepackage{graphicx}
\usepackage{bm}
\usepackage{amsmath,amssymb,amsthm}
\usepackage[colorlinks=true,linkcolor=blue,citecolor=red]{hyperref}
\newcommand{\be}{\begin{equation}}
\newcommand{\ee}{\end{equation}}
\newcommand{\bea}{\setlength\arraycolsep{2pt} \begin{eqnarray}}
\newcommand{\eea}{\end{eqnarray}}

\def\0{{\sst{(0)}}}
\def\1{{\sst{(1)}}}
\def\2{{\sst{(2)}}}
\def\3{{\sst{(3)}}}
\def\4{{\sst{(4)}}}
\def\5{{\sst{(5)}}}
\def\6{{\sst{(6)}}}
\def\7{{\sst{(7)}}}
\def\8{{\sst{(8)}}}
\def\sst#1{{\scriptscriptstyle #1}}

\makeatletter \@addtoreset{equation}{section}

\usepackage{multirow}
\definecolor{lime}{HTML}{A6CE39}
\newcommand{\orcidicon}{%
    \begin{tikzpicture}
    \draw[lime, fill=lime] (0,0)
        circle [radius=0.16]
        node[white] {{\fontfamily{qag}\selectfont \tiny ID}};
    \draw[white, fill=white] (-0.0625,0.095)
        circle [radius=0.007];
    \end{tikzpicture}   \hspace{-2mm}
}
\newcommand\orcidAdil{{\href{https://orcid.org/0000-0001-7623-5541}{\orcidicon}}}

\newcommand\orcidMohamed{{\href{https://orcid.org/0000-0003-1185-0062}{\orcidicon}}}

\begin{document}
%

\title{\normalsize
{\bf \Large	 Superentropic Black Hole Shadows  in Arbitrary  Dimensions   }}
\author{ \small   A.  Belhaj,  \orcidAdil\!\! \footnote{a-belhaj@um5r.ac.ma},  M. Benali\orcidMohamed\!\!\footnote{mohamed\_benali4@um5.ac.ma},  Y. Hassouni \footnote{ Authors in alphabetical order.}
	\hspace*{-8pt} \\
	{\small D\'{e}partement de Physique, Equipe des Sciences de la mati\`ere et du rayonnement, ESMaR}\\
{\small   Facult\'e des Sciences, Universit\'e Mohammed V de Rabat, Rabat,  Morocco} } \maketitle

 \maketitle
\begin{abstract}
We investigate  the shadow  behaviors  of the superentropic  black holes  in arbitrary dimensions.   Using  the   Hamilton-Jacobi  mechanism, we first  obtain   the associated null geodesic  equations of motion.  By help of   a spheric  stereographic projection, we  discuss    the shadows   in terms of  one-dimensional real curves.  Fixing the mass parameter $m$,   we  obtain  certain shapes being remarkably  different than four dimensional  geometric configurations. We  then study  theirs behaviors   by varying  the black hole  mass parameter.   We show that  the shadows undergo  certain geometric  transitions depending on  the spacetime  dimension.  In terms of  a critical  value  $m_c$, we find   that the four dimensional shadows  exhibit  three configurations   being  the   D-shape, the  cardioid and  the naked singularity   associated with $m>m_c$, $m=m_c$ and $m<m_c$, respectively.  We reveal  that  the D-shape passes to  the naked singularity via  a critical curve called  cardioid.  In  higher dimensions,  however,  we  show that  such  transitional behaviors  are removed.

	\end{abstract}
\newpage

\tableofcontents

\newpage

\section{Introduction}
Black hole physics   has received  a remarkable  interest   from many years. This physics  has become  primordial  to understand quantum gravity models.   The associated  contributions   have been supported   by the  gravitational wave detections and the  black  hole imaging provided by   Event Horizon Telescope  international collaborations  \cite{1,2,3}.  Concretely,  many   works have been elaborated dealing with   the  thermodynamic and  the optical aspects of such fascinating  objets.  Interpreting the pressure  as a cosmological constant in  Anti-de Sitter (AdS) geometries,   the   black hole thermodynamic  has  taken  a central place in gravity model investigations.  This provides new developments   in such a physics by   unveiling data on certain  transitions shearing similarities with Van der Waals fluids.  Precisely, the Hawking-Page transition has  been examined  for four and    higher dimensional gravity  theories  \cite{5,6,7}.   Precisely, it  has been  revealed   that such a transition   generates certain universalities   \cite{8,9}.  Moreover,  the optical aspect has  been  approached    by  investigating   the  deflection angle of  the light rays  and the  shadow  behaviors\cite{99,10,11,12,13, ca1,ca2,ca6,ca7,ca9,ca10,ca11,ca12 }.  In four dimensions, the black hole shadows  of various  black holes have  been engineered using  one dimensional real curves \cite{14,15,16,17}.  In particular, the  visualization of  the shadow casts  are  obtained from  the null geodesic equations. This finding    has been   supported  by the study of geometrical observables showing    information about the involved size and the shape of  such closed real curves. For non-rotating black holes, it has been revealed  that the  black hole shadows  exhibit  circular geometric configurations. It has been remarked that the associated  size can be controlled by internal and external  moduli spaces including  the dark field sector  \cite{10}.  The circular  geometric manifestation   can be distorted by introducing the   rotating  parameter which    generates   non-trivial  geometries  involving  either  D  or    cardioid  shapes  \cite{21,14,15,17,18,20}.  The latter has been  appeared in the study of  the AdS black holes  obtained from type superstrings and M-theory scenarios  using brane physics  \cite{20,200}.  Certain distorted  geometrical behaviors have been observed  for   rotating  stringy solutions exhibiting cardioid shapes by varying the brane number. \par
Most   recently,     the  pulsar SGR J174-2900 near  supermassive black holes  SgrA* has been investigated    providing   physical      aspects  of  the involved  horizon and the  horizonless  of events \cite{22,23}.     The relation between the thermodynamical volume aspect and the black hole entropy, including the associated area,  has  been investigated by the help of   the Reverse Isoperimetric Inequality\cite{25}. Concretely,  the black holes in  the  (AdS) spacetime provide interesting results. For generic  values of  the cosmological constant $\Lambda$,  the domaine of outer communication is bounded by a cosmological horizon. This  has been considered  as a relevant  relation which has been exploited to  link the optical  proprieties  of  the black hole with  the associated spacetime.
A special  interest  has been devoted   to   four dimensional superentropic  black holes   being a fascinate  solution  with non compact horizon topologies with  exceeding maximum bound  entropies  \cite{230,24,25}.

    Up to certain   limits, the superentropic  black holes   have  been  investigated by using an   ultra-spinning	limit of the  Kerr-Newman-AdS solutions \cite{26,27,28}.    The associated optical and thermodynamic  aspects have been studied.  Precisely, the thermodynamic  behaviors   of such black holes have been  examined     by   exploiting   ultra-spinning approximation limits \cite{29,30}.     Concerning the optical aspect,   the  four dimensional   shadows   have been  studied  using the  Hamilton-Jacobi  formalism \cite{14}.  Among others,   it has been found   many geometrical configurations  including ellipse shaped and  naked singularity behaviors.\par
Motivated by  various activities including  the optical  properties in  higher dimensional supergravity models \cite{99,230}, the   AdS space could open  many  interesting roads since it opens   windows  associated with the gauge-gravity duality in string theory and related topics.    Among others,   it has been remarked that the involved size parameter  $\ell$   has been liked to the mass parameter of the superentropic black hole and  and its  charge $Q$. In this way, the mass behaves differently.    This could bring  different  optical behaviors  compared  to the trivial black hole solutions.  Moreover,   the  intrinsic symmetries of the  associated metric  gives  a complete integrability  of the geodesic motion including  the separation of the Hamilton-Jacobi equations.  Constrains on such  black holes  could make contact with  current or future observations given by EHT collaborations.\\
 The aim of this work is to  investigate  the shadows    of  the superentropic  black holes  in arbitrary dimensions.  We  obtain   the corresponding null geodesic  equations of motion   using  the   Hamilton-Jacobi  scenarios.    Exploiting  a  spheric stereographic projection, we  approach   the shadows   in terms of  one-dimensional real curves.  Fixing the mass parameter,   we  find  certain shadow  shapes being remarkably  different than four dimensional  geometric configurations. We  then  examine   theirs behaviors   by varying  the black hole  mass parameter.    Concretely,  we  reveal  that  the shadows undergo transitions depending on  the spacetime  dimension.   Varying the mass  with respect to a critical  value  $m_c$, we  show    that the four dimensional shadows  exhibit  three configurations   being  the   D-shape,  the cardioid and  the naked singularity   associated with $m>m_c$, $m=m_c$ and $m<m_c$, respectively.   Precisely,  we  observe  that  the D-shape passes to  the naked singularity via the cardioid critical curve.  In  higher dimensions,   however,  we reveal  that such behaviors  are removed.\par
This paper is organized  as follows.  In section 2, we present  a concise review on superentropic black holes in higher dimensions.  In section 3,  we investigate the shadow behaviors in arbitrary dimensions.   In section 4, we reconsider the study of four dimensions by showing a possible geometric transitions in shadow behaviors.   In section 5, we provide a  study  for dimensions more than four. The last  section is devoted to  concluding  discussions. 
\section{Superentropic black holes in higher dimensions}
We start by exposing  a concise review on higher dimensional superentropic  neutral black hole solutions.  Certain physical aspects of these solutions have been dealt with  in arbitrary dimensions  \cite{27}.   They  could be considered as   new solutions  to the Einstien-Maxwell equations   being  supported by  supergravity models  relying on  extra dimensions.  In particular, we consider single rotating black hole solutions.  Following \cite{27,29},  the  associated metric  line element  corresponding to   a   $d$ dimensional spacetime   reads as 
\begin{eqnarray}
ds^{2}=&-&\frac{\Delta}{\rho^2} \left(dt - {\ell} \sin^2\theta d\phi \right)^2 +\rho^2  \left(\frac{dr^2}{\Delta}
+\frac{d\theta^2}{\sin^2\theta} \right)\\
\nonumber
&+& \frac{ \sin^4\theta}{\rho^2} \left( \ell dt -{(r^2+\ell^2)}d\phi\right)^2+ r^2\cos^2\theta d\Omega^2_{d-4},
\end{eqnarray}
where $\ell$ is a length parameter  linked to the  cosmological  constant. $\Delta$ and   $\rho^2$  are relevant functions  taking  the following form
\begin{equation}
\rho^2=r^2+\ell^2\cos^2\theta, \hspace{0.5 cm} \Delta=(\ell+\frac{r^2}{\ell})^2-2mr^{5-d},
\end{equation}
where  $m$  is  the mass  parameter.   $d\Omega^2_{d-4}$   
 denotes the line element of the $(d-4)$-dimensional unit sphere. To  obtain  a compact black hole object, one should introduce  a new chemical potential $K$  in order to  consider a periodic    direction    $\phi$   as follows $\phi \sim \phi + \alpha $,  where  $\alpha$ is a   dimensionless parameter\cite{26,27,29}.  A close inspection    shows that the existence of the black hole horizon    depends on  the spacetime dimension  $d$ as well as   on the  involved  moduli space.   In  $d=4$, for instance,  the  horizon existence   generates   a  constraint  between  the     black hole parameters  including $m$ and  $\ell$. It is given by  
\begin{equation}
\label{ES }
m\geqslant \frac{8}{3\sqrt{3}}\ell,
\end{equation}
where  the   critical mass parameter
\begin{equation}
m_c= \frac{8}{3\sqrt{3}}\ell, 
\end{equation}
will  be   involved  in the discussion of the  shadow behaviors of such  black holes.  For  $d\geqslant5$, however,  the above constraint  reduces to a simple  one   provided by $m>0$.  
Having elaborated the essential backgrounds, we move to investigate the  superentropic  neutral  black  hole shadow behaviors  in  higher dimensions.  Varying the mass parameter,  we will show that  the shadow   geometries undergo certain  geometric transitions.  This could  be interpreted as  possible transitions in the  optical aspect  going beyond to  the ones observed in thermodynamics. This feature could be illustrated  in terms of one dimensional  real curves embedded  in a  two-dimensional plane supported by the above metric form. 

\section{Shadows   in arbitrary dimensions}
\begin{figure}[!ht]
\centering
\includegraphics[scale=.4]{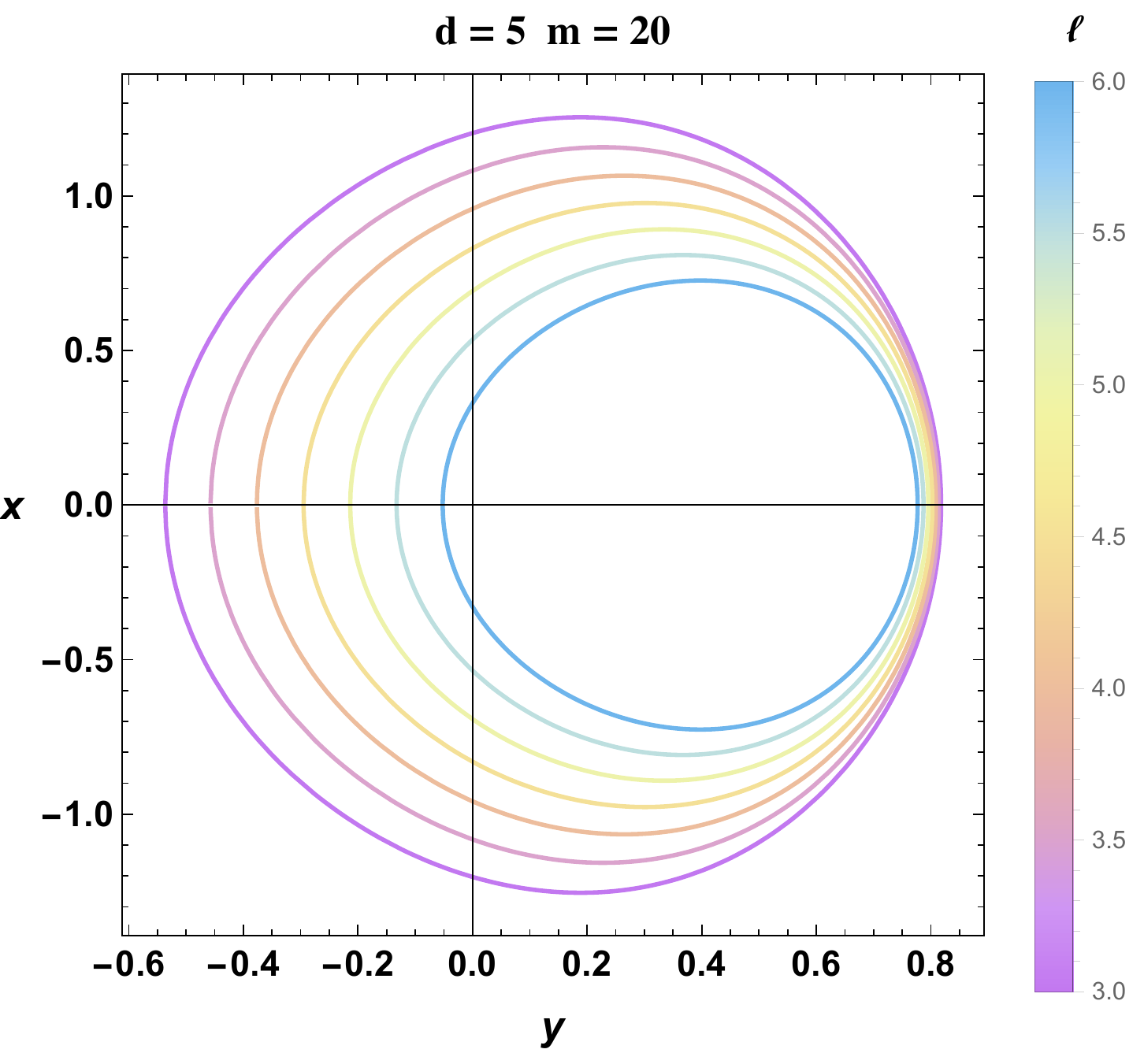} 
\hspace{2 cm}
\includegraphics[scale=.49]{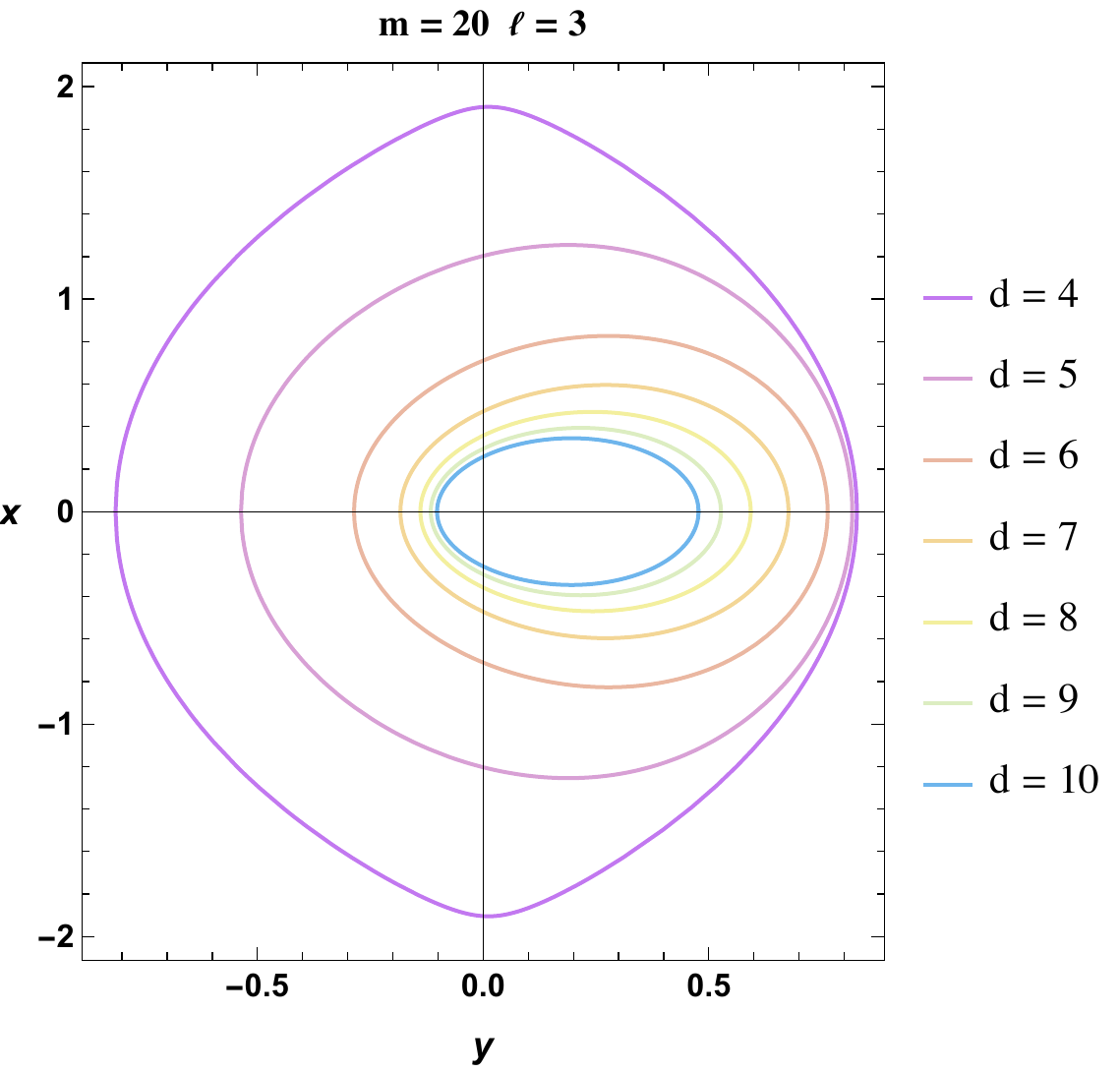} 
\caption{{\it \footnotesize {Shadows  of superentropic  black holes. Left panel:  Five dimensional shadows  by varying $\ell$ and fixing $m$.  Right panel:  Shadows for different values of $d$ by fixing    $m$ and  $\ell$. The observer is positioned at $r_{ob}=50$ and $\theta_{ob}=\frac{\pi}{2}$.}}}
\label{Rb}
\end{figure}

Motivated by string theory and  related supergravity models,  we  would like   to  study the  shadow  behaviors of   superentropic  neutral  black  holes in arbitrary dimensions.
Before   elaborating  shadow geometries in arbitrary dimensions, we  establish first   the null  geodesic equations  of motion.    Employing  the Hamilton-Jacobi  formalism, we  write down    the equations of  the photons near  the  superentropic black hole  horizons.  Following  \cite{31},    certain relations    are needed. Indeed,  one has 
\begin{eqnarray}
\label{ks1}
0&=&\frac{\partial S}{\partial \tau}+\frac{1}{2}g^{\mu\nu}\frac{\partial S}{\partial x^\mu}\frac{\partial S}{\partial x^\nu}
\end{eqnarray}
where  $\tau$   is   an affine parameter along the geodesics.  The action  $S$ is proposed  to take the following  form 
\begin{eqnarray}
\label{ks2}
S&=&-Et+L_\phi\phi+S_r(r)+S_{\theta}(\theta)+\sum_{i=1}^{d-4}S_{\psi_i}(\psi_{i})
\end{eqnarray}
where one  has used $E=-p_t$ and  $L=p_\phi$   being the total energy and the  angular momentum of the photons, respectively.   In this regards,    $S_r(r)$, $S_{\theta}(\theta)$ and    $S_{\psi_{i}}(\psi_{i})$ represent     functions depending on $r$, $\theta$ and $\psi_{i}$ variables, respectively.  It is worth noting that the variables  $\psi_{i}$ and the   functions $S_{\psi_{i}}(\psi_{i})$  are associated with  the extra dimensions.   Sending these extra dimensional functions to zero, we recover the  expressions of the four dimensional action  reported   in  \cite{14}. Using  the separation method and the Carter constant, we  can  get   the complete null geodesic equations.  Precisely,  they are  given by  \begin{align}
\rho^2 \frac{d  \, t}{d \tau}& =  E \left[ \frac{\lambda\left(r^2 +\ell^2 \right) }{\Delta} +  \frac{ \ell \left(\xi  - \ell \sin^2 \theta \right) }{ \sin^2\theta} \right], \\
\label{r}
\rho^2  \frac{d  \, r}{d \tau} &=\sqrt{\mathcal{R}(r}),\\
 \rho^2\frac{d  \, \theta}{d \tau} & =\sqrt{\Theta(\theta)}\\
 \label{phi}
\rho^2 \frac{d  \, \phi}{d \tau} &= E \left[ \frac{\lambda\;\ell}{\Delta} + \frac{ \xi-\ell \sin^2 \theta}{\sin^4 \theta }  \right],\\
  \rho^2\frac{d  \, \psi_i}{d \tau} & =\sqrt{\Psi_i(\psi_i)},  \qquad  i=1,\ldots,d-4
\end{align}
where   one has used    $\lambda=  \left(r^2 +\ell^2 \right)- \ell \xi$.    The quantities $\xi=\frac{L_{\phi}}{E}$ and $\eta=\frac{\mathcal{K}}{E^2}$,  representing the  impact parameters,  have been introduced in such equations. $\mathcal{K}$ is  a  separable constant\cite{31,10}.      The computation shows  that   $\mathcal{R}(r)$,   $\Theta(\theta)$,  and  the extra dimension functions $\Psi_i(\psi_i)$  take the following forms 
\begin{eqnarray}
\mathcal{R}(r)  = &E^2&\left[ \lambda^2- \Delta\left({(\ell-\xi)^2}+\eta \right)  \right],\\
\Theta(\theta)  =  &E^2& \left[ \eta \sin^2\theta  - \xi^2\cos^2\theta\left(\cot^2\theta-2( \frac{\ell}{\xi}-1)\right ) \right],\\
\Psi_i(\psi_{i}) & = & E^2\left[2\eta\rho^2\sum_{k=1}^i r^2\cos^2\theta\prod_{j=1}^{k-1}\sin^2\psi_j \right].
\end{eqnarray}
 At this level, it is  interesting  to  comment these equations. 
Taking $\psi_{i}=0$,   we recover the  four dimensional   geodesic equations reported in  \cite{14}.  The  functions  $\Psi_i(\psi_{i})$ and   $\Delta$   share data on the shadow  behaviors in higher dimensions. The   radial and the polar   contributions in   $\Psi_i$ could be understood in   terms of  the fibration  properties used in the compactification scenarios of higher dimensional supergravity models including  superstrings,  M-theory, and related topics.  This  means that four dimensional models   could  be considered as a   base space   where the  ($d-4$)-dimensional  real sphere  moves on it. 
 Roughly,  the   unstable circular of  the photons around the black hole horizon  can be obtained  by solving the following equations
\begin{equation}
\mathcal{R}(r)\Big|_{r=r_s}=\frac{d\, \mathcal{R}(r)}{d r}\Big|_{r=r_s}=0,
\end{equation}
where $r_s$ is the circular orbit radius of the photon\cite{10,16,17,18}.  The computations provide
\begin{align}
& \eta =\frac{r^2 \left(16 \ell^2 \Delta-\csc ^2\theta  \left(4 \Delta-r \Delta^{\prime}\right)^2\right)}{\ell^2{ \Delta^{\prime}}^2}\bigg\vert_{r=r_s},\\
& \xi=\frac{\left(r^2+ \ell^2\right) \Delta^\prime-4r \Delta }{\ell \Delta^\prime} \bigg\vert_{r=r_s}.
\end{align}
Fixing  the observer  distance  $r_{ob}$, we can find   the shadow  behaviors    in  the domain of  outer communications  $(\Delta>0)$\cite{15,16,17}. In this way,  the corresponding   vectors of  the observer  are needed   to get the associated  null geodesic equations of motion.   The extra dimensions    push one to introduce  new  vectors   from four dimensional point  of views. These vectors are given by 
\begin{eqnarray}
\label{e_0}
e_0 & = & \frac{ (r^2+\ell^2)\partial_t+\ell\partial_\phi}{\sqrt{\Delta \rho^2}} \bigg\vert_{(r_{ob}\,,\theta_{ob})},\\
\label{e_1}
e_1 & = & \frac{\sin\theta}{\sqrt{\rho^2}} \partial_{\theta}\bigg\vert_{(r_{ob}\,,\theta_{ob})},\\
\label{e_2}
e_2 & = &  -\frac{ \ell\sin^2{\theta}\partial_t+\partial_\phi}{\sqrt{\rho^2}\sin^2{\theta}} \bigg\vert_{(r_{ob}\,,\theta_{ob})}, \\
\label{e_3}
e_3 & = &-\frac{\sqrt{\Delta}}{\sqrt{\rho^2}} \partial_r\bigg\vert_{(r_{ob}\,,\theta_{ob})} \hspace{0.2 cm},\\
\label{e_i}
e_{i+3} & = & \frac{1}{\sqrt{\sum\limits_{k=1}^{i-3} r^2\cos^2\theta\prod\limits_{j=1}^{k-1}\sin^2\psi_j}} \partial_{\theta_i} \bigg\vert_{(r_{ob}\,,\theta_{ob})}
\end{eqnarray}
where one has used  $i=1,\ldots,d-4$. The timelike vector $e_0$  indicates    the four-velocity of the  observer and  $e_3$   represents   the  third vector along the spatial direction pointing toward the center of the black hole.  Here,   $e_0\pm e_3$   are considered  as tangent  directions to the  one  of principal null congruences where  $r_{ob}$ and $\theta_{ob}$  are the   distance and the angle of  the observer, respectively.  The vectors $e_{i+3}$  are associated with the higher extra  dimensions. Taking $d=4$ and evincing such vectors, we recover the four dimensional ones    proposed in  \cite{14}. In  generic configurations,  the   light equation  being   tangent  to the observer  position   can be defined via  the  relation
 \begin{equation}
\label{lambdat}
\dot{\lambda}=\dot{r}\partial_r+\dot{\theta}\partial_\theta+\dot{\phi}\partial_\phi+\dot{t}\partial_t+\sum_{i=1}^{d-4}\dot{\psi_i}\partial_{\psi_i} 
\end{equation}
where the $(d-1)$ vectors of the spacelike   can be represented  in a basis corresponding to the  spherical coordinates. This  can be exploited to   establish  the tangent equation  in terms  of the orthonormal vectors $\{e_0,\ldots,e_{d-1}\}$ and  the celestial coordinates $(\gamma,\delta, \sigma_i)$.   Using the spherical coordinates in higher dimensions, we obtain 
\begin{equation}
\begin{split}
\label{lambdatt}
&\dot{\lambda}=\beta\Big(-e_0+\cos\delta e_3+\sin\delta \cos\gamma e_1+\sin\delta \sin\gamma \cos\sigma_1 e_2+\\
&\sin\delta \sin\gamma\left[\sum_{i=1}^{d-5}\left(\prod_{j=1}^{i-1}\sin\sigma_{j}\right)\cos\sigma_{i+1}e_{i+3}+\prod_{i=1}^{d-4}\sin\sigma_ie_{d-1}\right]\Big)
\end{split}
\end{equation}
where $\beta$ is  a scalar factor. Combining  the  equations of the  light  rays and Eq.(\ref{lambdatt}),   we get 
\begin{equation}
\label{alpha}
\beta=g(\dot{\lambda},e_0)=\frac{E}{\sqrt{\Delta_r\rho^2}}\Big(\ell\xi-(r^2+\ell^2)\Big) \bigg\vert_{(r_{ob}\,,\theta_{ob})}.
\end{equation}
An examination reveals that  the celestial coordinates   are   functions  only of the impact  parameters $\xi$ and $\eta$ needed  to illustrate   one-dimensional  real curves describing  the associated shadow behaviors.  Exploiting  the  null geodesic equations and  comparing   the coefficients  of  Eq.(\ref{lambdat})  with  Eq.(\ref{lambdatt}), we can express the  celestial coordinates  in terms  of  $\xi$ and $\eta$.  Indeed, it follows  that the spheric coordinates should verify 
\begin{eqnarray}
\label{rho}
\sin{\delta}&=&\frac{\pm\sqrt{\Delta\eta}}{((r^2+\ell^2)-\ell\xi)}\bigg\vert_{(r_{ob}\,,\theta_{ob})},\\
\label{psi}
\sin{\gamma}&=&\frac{\sqrt{\Delta}}{\sin\delta}\left(\frac{(\ell-\csc^2{\theta}\xi)}{\ell\xi-(r^2+\ell^2)}\right)\bigg\vert_{(r_{ob}\,,\theta_{ob})},\\
\sin\sigma_i&=& \frac{r^2\cos^2\theta\sqrt{2\eta\Delta}}{\sin\delta \sin\gamma\prod\limits_{j=1}^{i-1}\sin\sigma_j}\left(\frac{\sum\limits_{k=1}^{i}\prod\limits_{j=1}^{k-1}\sin^2\psi_j}{\ell\xi-(r^2+\ell^2)}\right)\bigg\vert_{(r_{ob}\,, \theta_{ob})}.
\end{eqnarray} Applying  the  $\mathbb{S}^{d-2}$ sphere stereographic projections, we  can    get   the local  cartesian coordinates of  the coordinate  system ($x,y, z_1, \ldots, z_{d-4}$).  These coordinates   could be exploited     to represent the shadow geometries in  an appropriate   spheric projection.  The computations give 
\begin{equation}
\label{x}
\begin{split}
x = & -2 \tan{(\frac{\delta}{2})}\cos{\gamma},\\
y  = & -2 \tan{(\frac{\delta}{2})}\sin{\gamma}\cos{\sigma_1},\\
z_{j}  = & -2 \tan{(\frac{\delta}{2})}\sin{\gamma}\left(\prod_{k=1}^{d-5}\sin\sigma_{k}\right)\cos\sigma_{j+1}\\
z_{d-4} = & -2 \tan{(\frac{\delta}{2})}\sin{\gamma}\left(\prod_{k=1}^{d-5}\sin{\sigma_k}\right)\sin{\sigma_{d-4}}
\end{split}
\end{equation}
where  one has used $ j=1,\ldots ,d-5$. In extra dimensions,  it  has been remarked that  the equations needed to get  such geometries   describing the   optical aspects of    the neutral  superentropic black  holes  involve  a  factor given by  $\cos^2\theta$.  Placing  the observer  in the equatorial plane,  we get 
\begin{equation}
\label{ }
\sin\sigma_{i}=0   \hspace{0.5 cm}  i=1,\ldots,d-4.
\end{equation}
Up to the periodicity conditions, it  is obvious that these constraints   are  solved  by 
\begin{equation}
\label{ }
\sigma_{i}=0.
\end{equation}
These conditions automatically impose  $z_{i}=0, \;   i=1,\ldots,d-4$. It is interesting to  note that  the remaining  information on  the extra dimensions are now  hidden only in $\Delta$. This allows one to consider only  the cartesian coordinates   ($x,y$) to visualize the shadow behaviors in arbitrary dimensions  by exploiting  one-dimensional real curves. This situation  of the   observer  matches perfectly  with   the   stereographic projection procedure  $
 \mathbb{S}^{d-2} \longmapsto \mathbb{R}^2$.
In this way,    the relevant parameter in the  shadow discussion  is  the   dimension $d$.  In what follows, we  inspect such optical behaviors by varying two essential parameters $m$ and $d$. First, we consider the variations  of the dimension $d$.  After that, the mass variation will be discussed.  In Fig(\ref{Rb}), we  illustrate  the shadow behaviors  for  different values of $d$ by fixing $m$ and varying  $\ell$. 
 The left panel  presents the shadow behaviors of  the special case $d=5$  for different values of  $\ell$.   It has been remarked that  the shadow size increases by decreasing $\ell$.    The right panel  shows the effect of the spacetime dimension $d$ on  the shadow aspect.  It has been observed that the  shadow size decreases by increasing $d$.   An examination reveals that  the   geometric configurations 
 of the  higher dimensional black hole solutions  are different than  the ones obtained in $d=4$  \cite{14}. In particular, the elliptic geometry has been modified.\\
   It should be interesting to make contact  with the shadow observation  of the supermassive black hole  associated with  $M87^\ast$ data,  obtained by the  EHT international collaboration. Indeed, the observational data can put certain  constraints on the relevant black hole parameters. Motivated by such activities,  we could    compare the shadow of  the superentropic black hole with  such observational data  by taking   $M = 1$  in units of the $M87^\ast$ black hole mass given by $M_{BH} = 6.5\times109\,M\odot$ and $r_0 = 91.2\,kpc$.
 According to\cite{1, 2 ,31aa}, it has been  remarked that  the experimental shadow size is around $5.19$.  However,   the shadow size of  the superentropic black hole is around  $2$ for generic regions of the moduli space.  The shadow size given by EHT is bigger   compared  with  the present studied  black holes.  We believe that such a difference is due to the involved  geometry. Indeed,  the  shadow shape  given   by  EHT  is almost  D-shape circle. However, the  shadow   shape of  the superentropic black holes    involve an  elliptic geometrical  form. This  could be supported by  the  relation between  the mass parameter  $m$ and $\ell$ of  the superentropic black holes.   For higher dimensional  theories,  we could speculate on  a possible link with primordial black holes having certain relations with extra dimension models.   This could be supported by the fact  that such black holes,   involving a  small mass parameter, exhibit also a   small length scale. We could expect that such black holes could  find  a place  in future observational  data associated  with  EHT collaborations including the recent one.      
 
  \section{Shadow transitions in four dimensions}
\begin{figure*}[!ht]
		\begin{center}
		\begin{tikzpicture}[scale=0.2,text centered]
		\hspace{-1.2 cm}
\node[draw, line width=1pt,color=orange,name=plan,dashed] at (-40,1){\small  \includegraphics[scale=0.40]{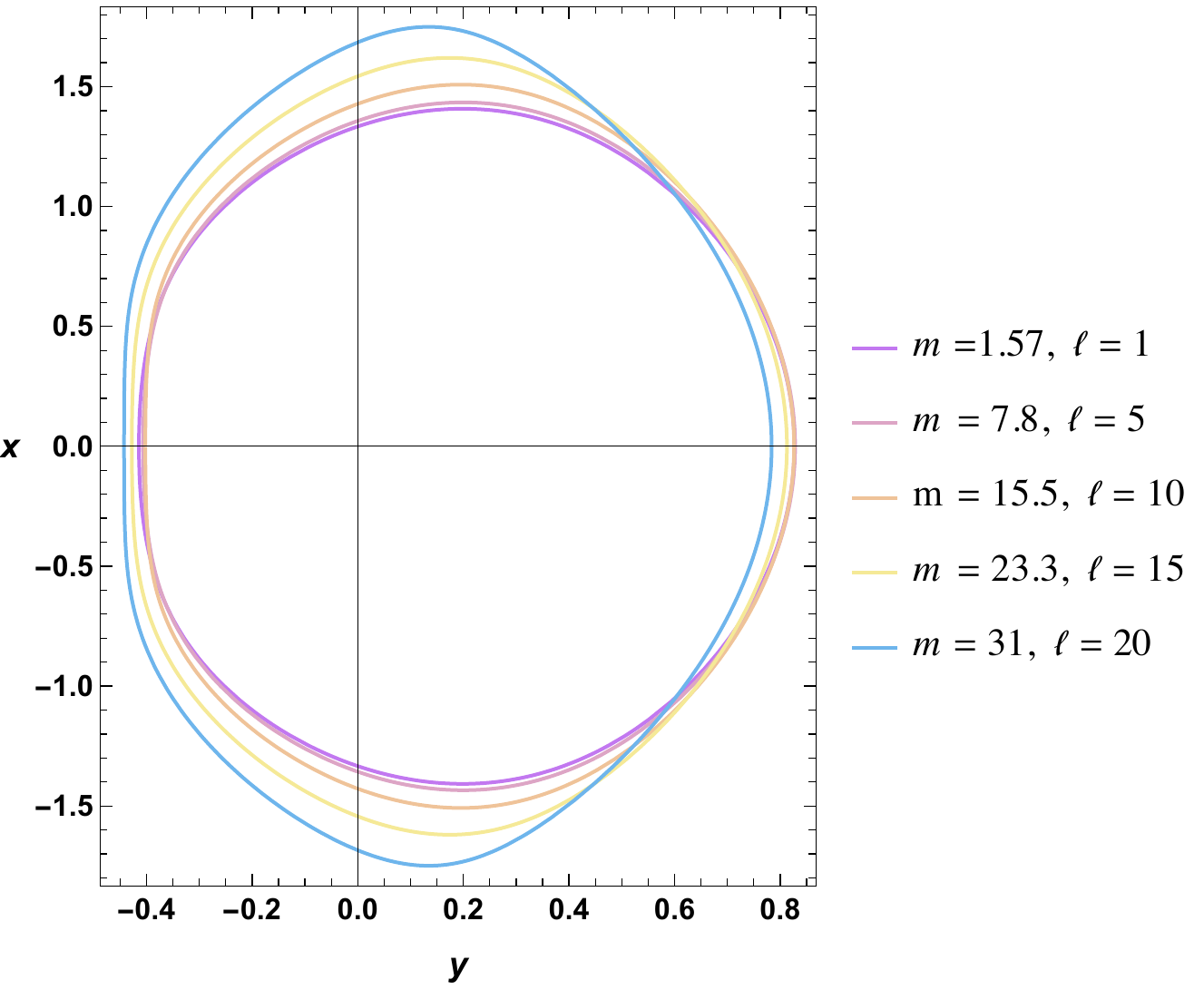}};
\node[draw, line width=1pt,color=green,name=plan,dashed] at (-10,1){\small  \includegraphics[scale=0.40]{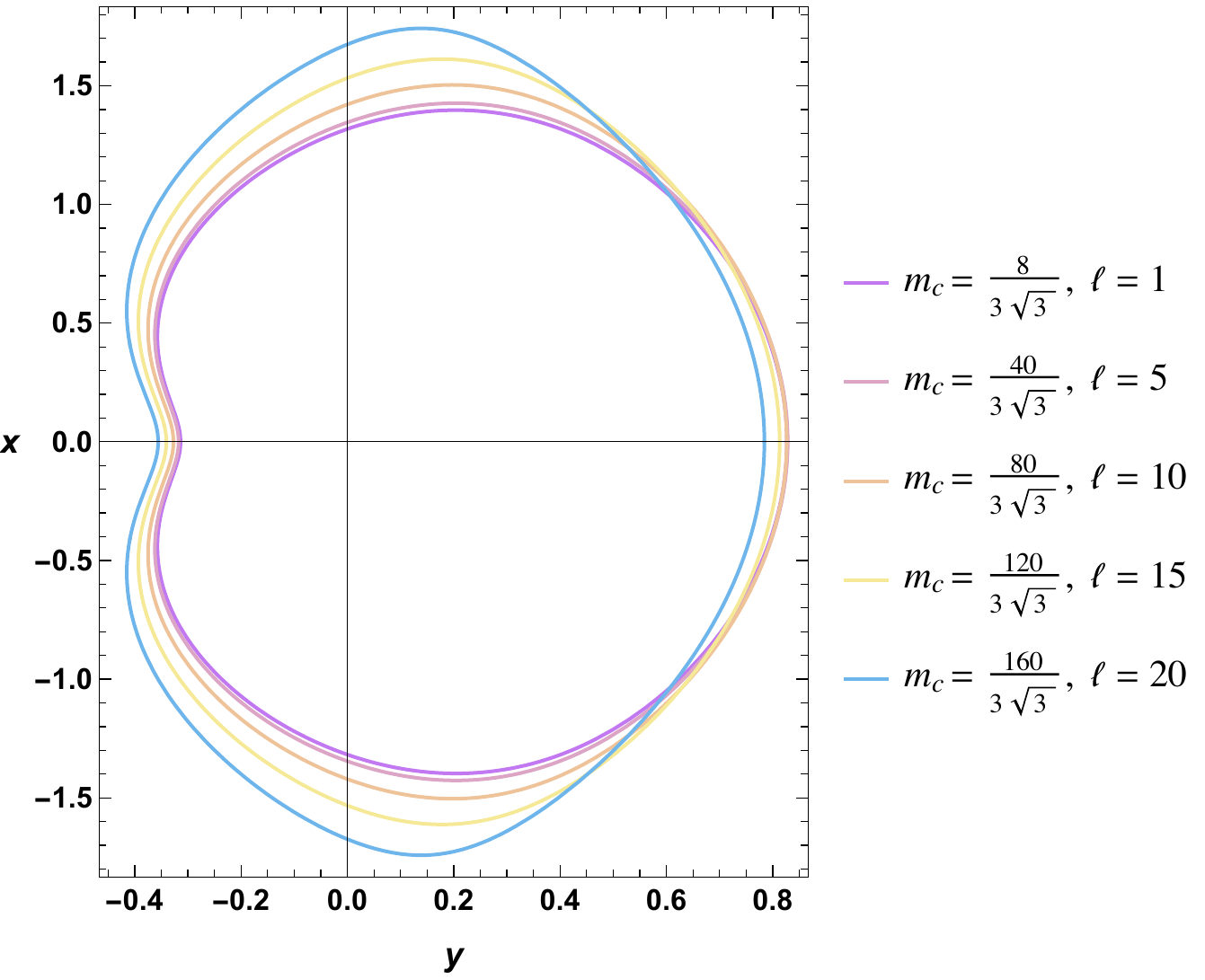}};
\node[draw, line width=1pt,color=purple,name=plan,dashed] at (20,1){\small  \includegraphics[scale=0.40]{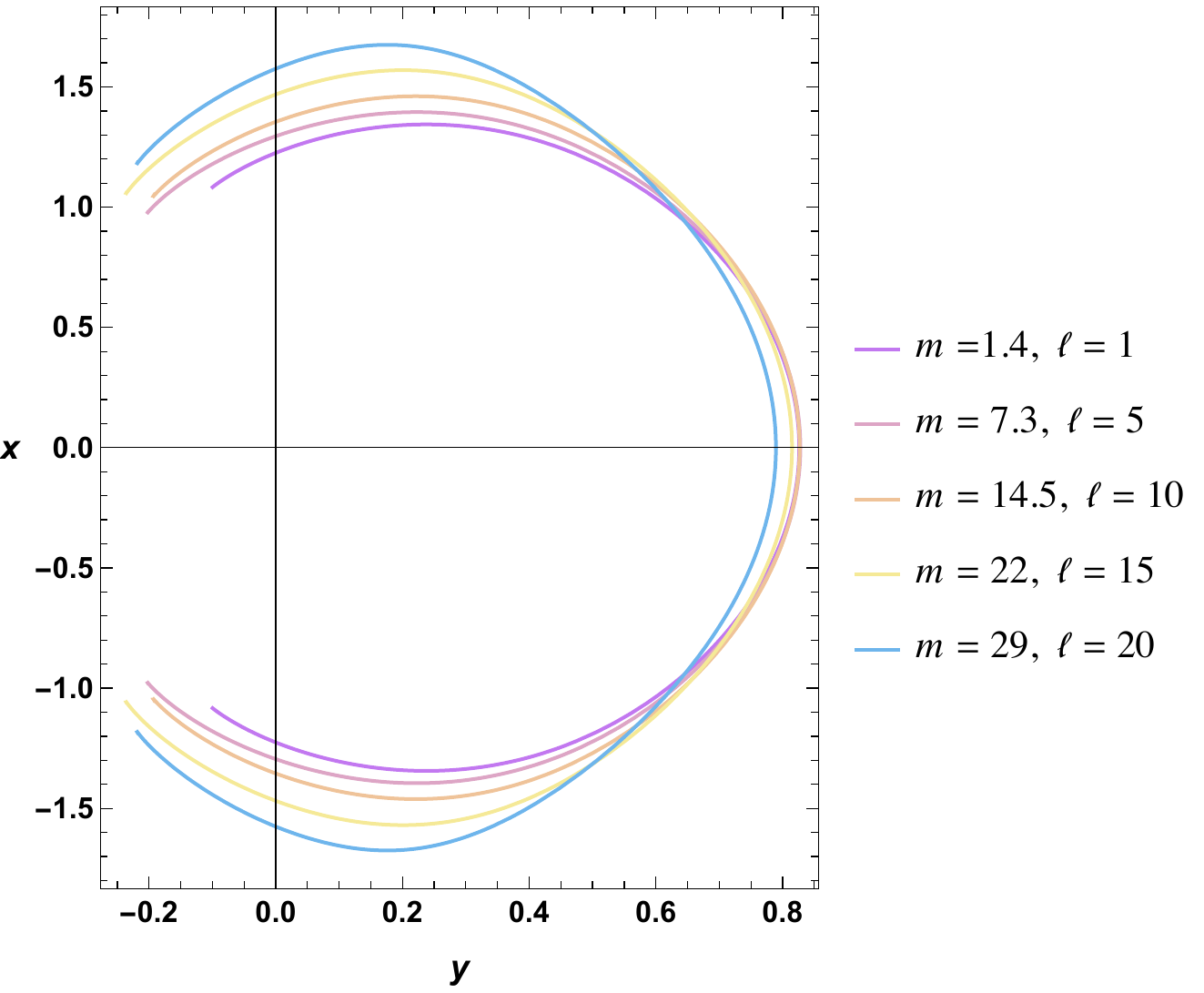}};	
\node[color=purple] at (20.5,-15.5) {Naked singularity};
\node[color=green] at (-10,-15.5) {Cardioid};
\node[color=orange] at (-40,-15.5) {D-shape};
\node[color=orange] at (-40,15.5) {$m> m_c$};
\node[color=green] at (-10,15.5) {$m= m_c$};
\node[color=purple] at (20.5,15.5) {$m< m_c$};
\end{tikzpicture}	
\caption{Shadow geometries of  four dimensional superentropic AdS black holes for different values of $m$ and $\ell$ in four dimensions. The observer is positioned at $r_{ob}=50$ and $\theta_{ob}=\frac{\pi}{2}$.}
\label{aa}
\end{center}
\end{figure*}
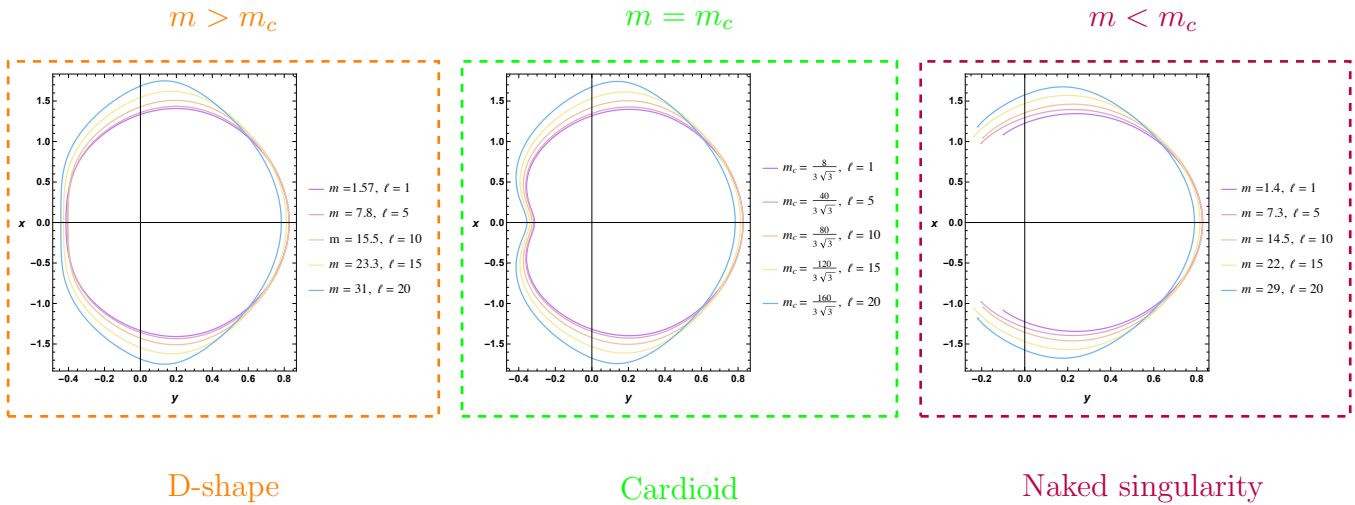

A close inspection on  the study of the black hole shadows  shows  that the   involved geometries   exhibit several configurations including  circular and D-shapes. In AdS backgrounds,  a non expected geometry called   cardioid have been found\cite{17,20,21}.   We will  show  that this geometry  is relevant to unveil a nice phenomena in shadow behaviors which could be understood as a transition in  the optical aspect of   the neutral  superentropic black holes in certain  dimensions.   Moreover, the    elliptic geometry  arises  naturally in  the domain of the horizon existence\cite{14,15,16,17}.  Due to the horizonless,  however, the naked singularity for  certain  black hole solutions  has   been observed  \cite{17,32,14}.  It is worth noting that   it  appears when $\Delta$   involves    complex roots.  Motivated by non-trivial horizon  geometries of  the  superentropic black holes, we  study  the associated  shadow behaviors   by varying the mass parameter being fixed in the previous  investigations. Various dimensions can  be dealt with.  In this way, the roots  of the  equation 
\begin{equation}
\label{deltafunction}
(\ell+\frac{r^2}{\ell})^2-2mr^{5-d}=0
\end{equation}
will be needed in the elaboration of   the  shadow behaviors.
The  geometry and  the mass constraints   of such black holes   push   one  to unveil  new  data on the associated  shadow aspects. To   show that, we  first   reconsider the study of  four dimensions.    Precisely, we  investigate  the shadow geometrical configurations  for   the superentropic black holes  by varying the mass parameter $m$   with respect to the critical  value   $m_c$  in certain  ranges of  the AdS  radius   $\ell$.   Three  situations  $m>m_c$,  $m=m_c$,   and  $m<m_c$  will be examined.   In Fig.(\ref{aa}),  we present the associated behaviors. 
It follows form this figure that the shadows  of these black holes  exhibit  an interesting phenomena according to the value of the critical  mass parameter.  It has been remarked  that for  $m<m_c$, we obtain  the naked singularity.   For $m>m_c$, however, the horizon of the black hole exists and the  corresponding shadow     involves    either the  elliptic or  the D-shape elliptic geometry for the mass  values  bigger or a litter bigger   to $m_c$, respectively, for    different   values of $\ell$.    Considering    a generic  value of  $\ell$ and  identifying $m$  with $m_c$, an unusual cardioid shadow  geometry appears.   The latter  could be supported by  limaçon approximations in which such   a geometry can be considered as  a special case.  It is noted  that this  approximation  provides   similar configurations using  other methods\cite{17,21}.    Varying the mass, these   three different   geometric configurations  can be  obtained  by passing   from one  shape  to another  one. We refer to them as a transition in  the optical aspect. To check  this phenomena, we consider  higher dimensional black holes.
\section{\bf{ Shadow behaviors of  higher dimensional  solutions}}
It seems possible to extend  the previous analysis  to higher dimensions. The extension of the  four dimensional transition picture to higher dimensions is based on Eq.(\ref{deltafunction}).  A rapid  examination shows that one should consider   two  situations  associated with $d=5$ and $d>5$, respectively.\\
\subsection{Five dimensional behaviors}
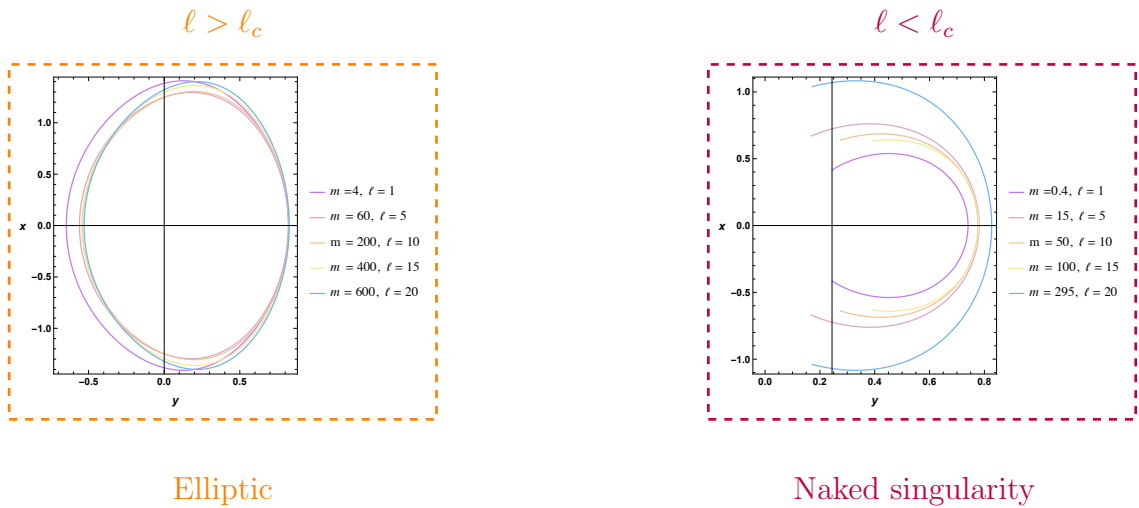
\begin{figure*}[!ht]
		\begin{center}
		\begin{tikzpicture}[scale=0.2,text centered] 
\node[draw, line width=1pt,color=orange,name=plan,dashed] at (-30,1){\small  \includegraphics[scale=0.4]{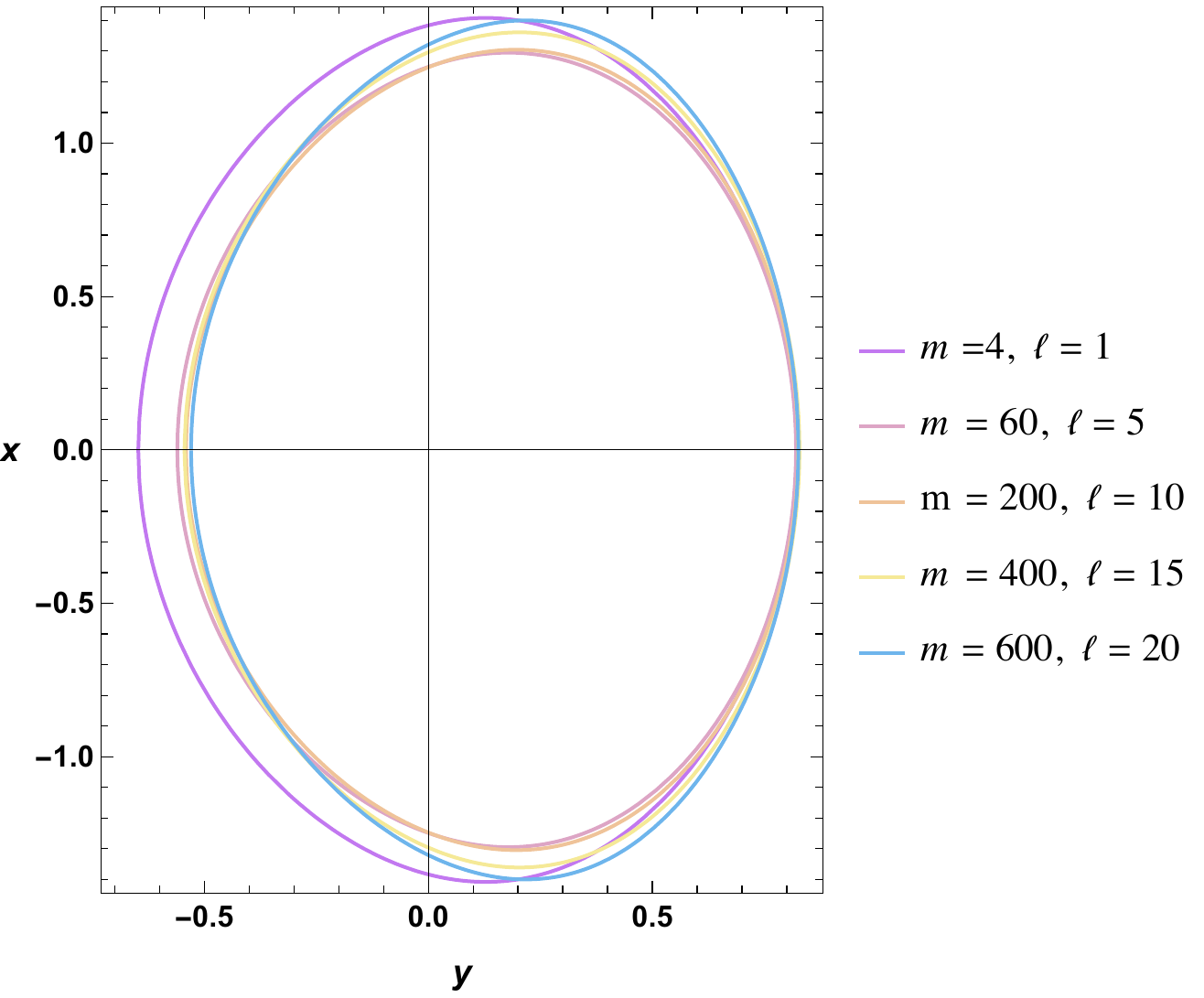}};
\node[draw, line width=1pt,color=purple,name=plan,dashed] at (16,1){\small  \includegraphics[scale=0.4]{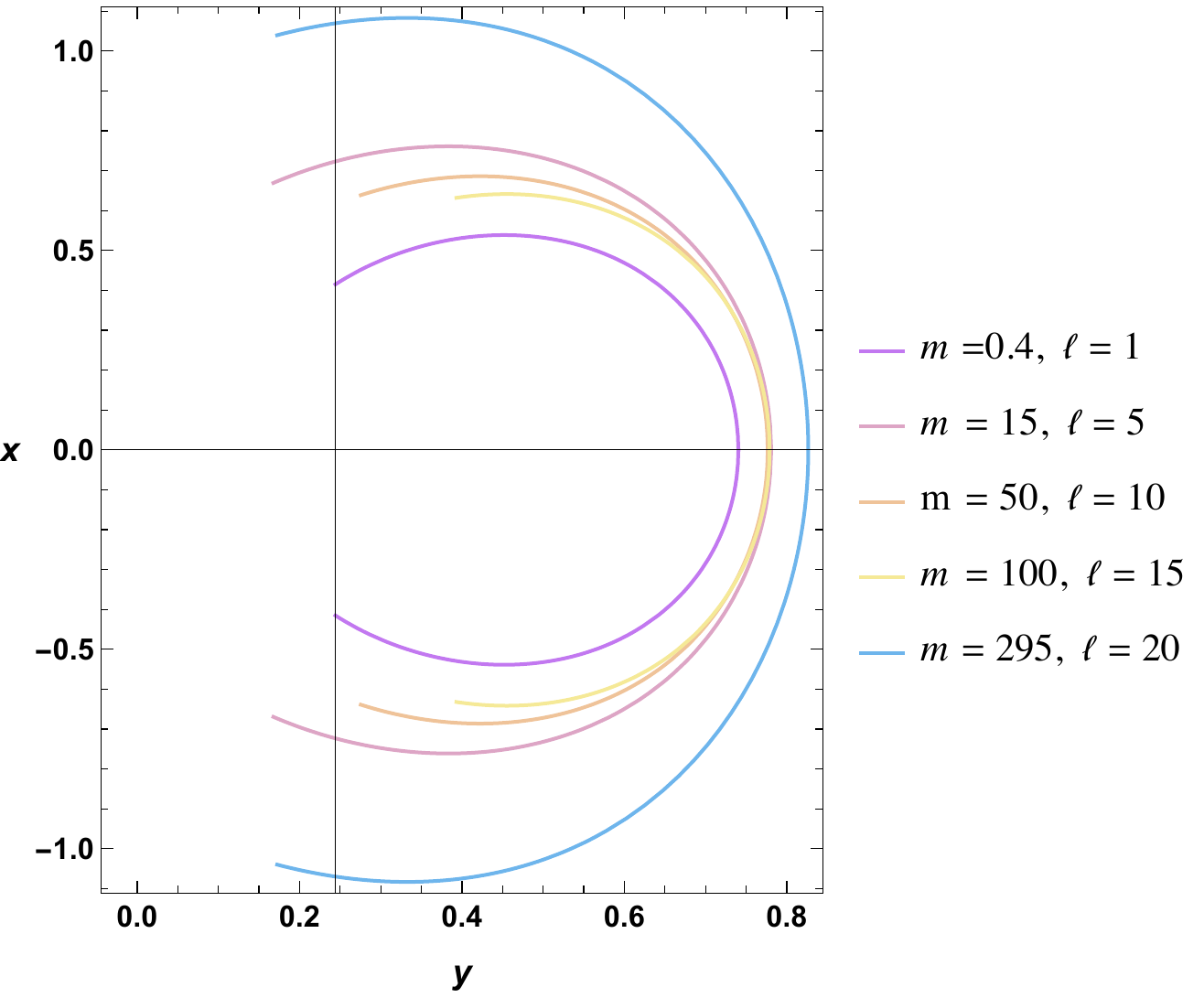}};	
\node[color=purple] at (15.5,-15.5) {Naked singularity};
\node[color=orange] at (-30,-15.5) {Elliptic};
\node[color=orange] at (-30,15.5) {$\ell> \ell_c$};
\node[color=purple] at (15.5,15.5) {$\ell< \ell_c$};
\end{tikzpicture}	
\caption{Shadow geometries of  the superentropic AdS black holes for different values of $m$ and $\ell$ in four dimensions. The observer is positioned at $r_{ob}=50$ and $\theta_{ob}=\frac{\pi}{2}$.}
\label{a1}
\end{center}
\end{figure*}

We first  engineer  the shadow shapes for five-dimensional solutions by using the above mentioned   stereographic projection.  Considering  the constraint  $m>0$    and taking  different values of $m$ and $\ell$, we  can approach the shadow behaviors.  Solving $\Delta=0$ in five dimensions, we   find  that one has only two geometric configurations based on the following constraints 
\begin{equation}
\label{ }
 \left\{
    \begin{array}{ll}
       \ell>\sqrt{2m} & \text{elliptic shape} \\
       &\\
         \ell<\sqrt{2m}& \text{naked singularity} .
    \end{array}
\right.
\end{equation}
The  associated  shadows are presented in   in Fig.(\ref{a1}).
It has been remarked that the cardioid geometry disappears.  In this way, the shadow geometry passes directly from  a  naked singularity to  an  elliptic geometry. An examination shows that these behaviors  are different  than   the four dimensional ones.   Fixing $\ell$, indeed,   the naked singularity  arises for small   mass  values   contrary to four dimensions in which large values are needed.  Similar aspects   are  observed  in elliptic geometries.\\
\subsection{Behaviors in more than five dimensions}
\begin{figure*}
		\begin{center}
		\begin{tikzpicture}[scale=0.2,text centered] 
		\hspace{-1.4 cm}
\node[] at (50,1){\small  \includegraphics[scale=0.4]{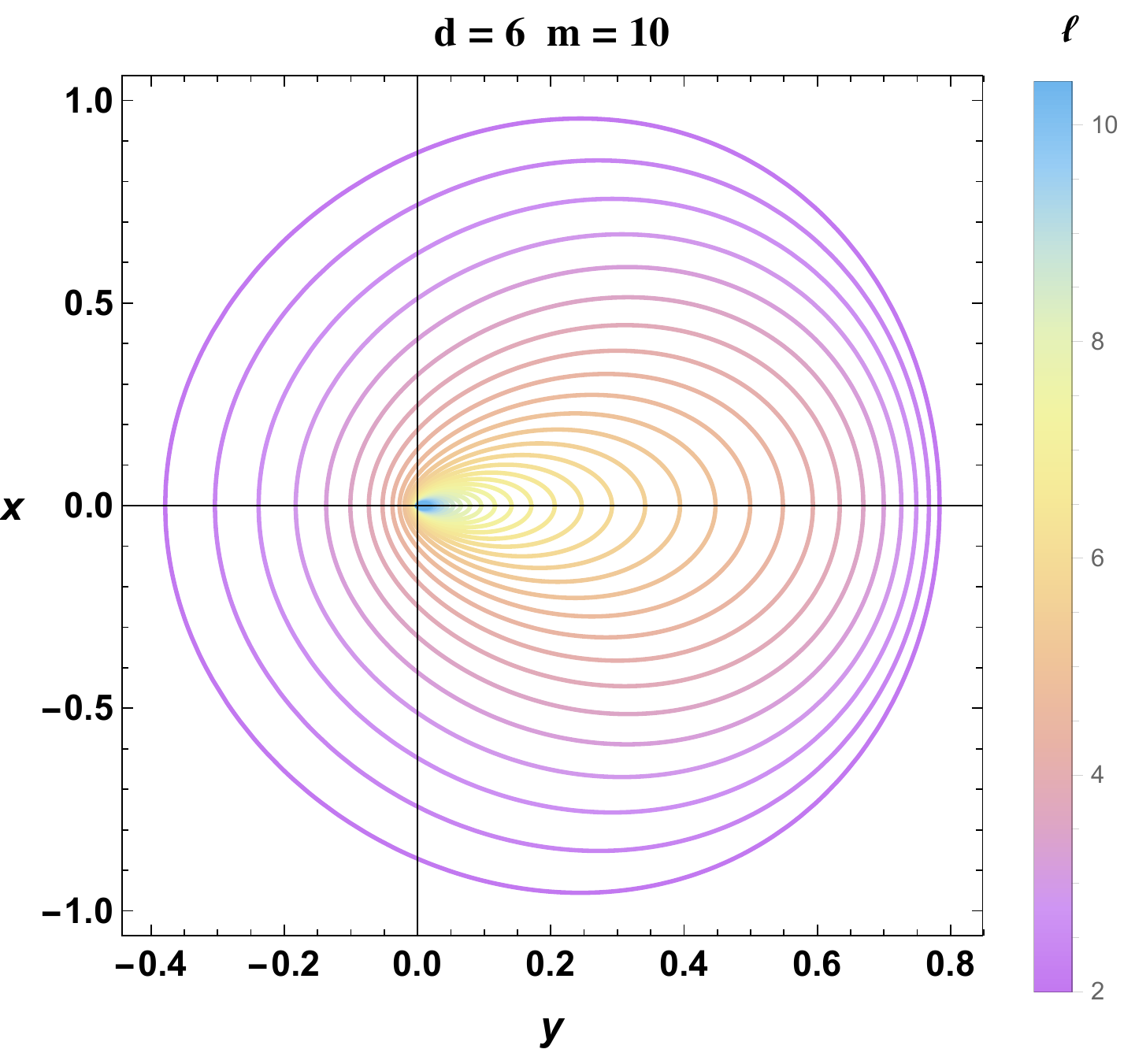}};
\node[draw, line width=1pt,color=red,name=plan,dashed] at (20,1){\small  \includegraphics[scale=0.3]{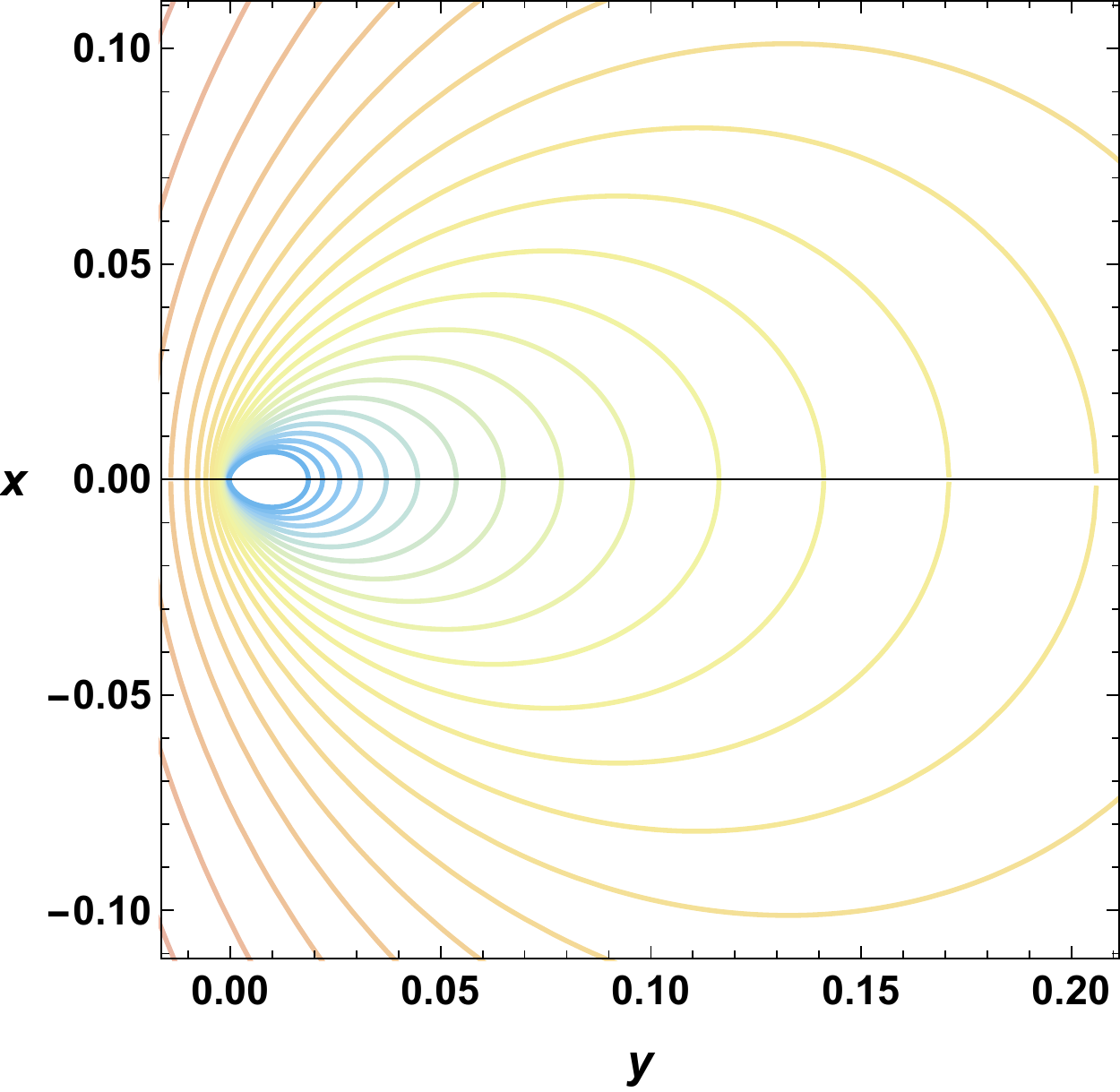}};
\node[] at (-10,1){\small  \includegraphics[scale=0.4]{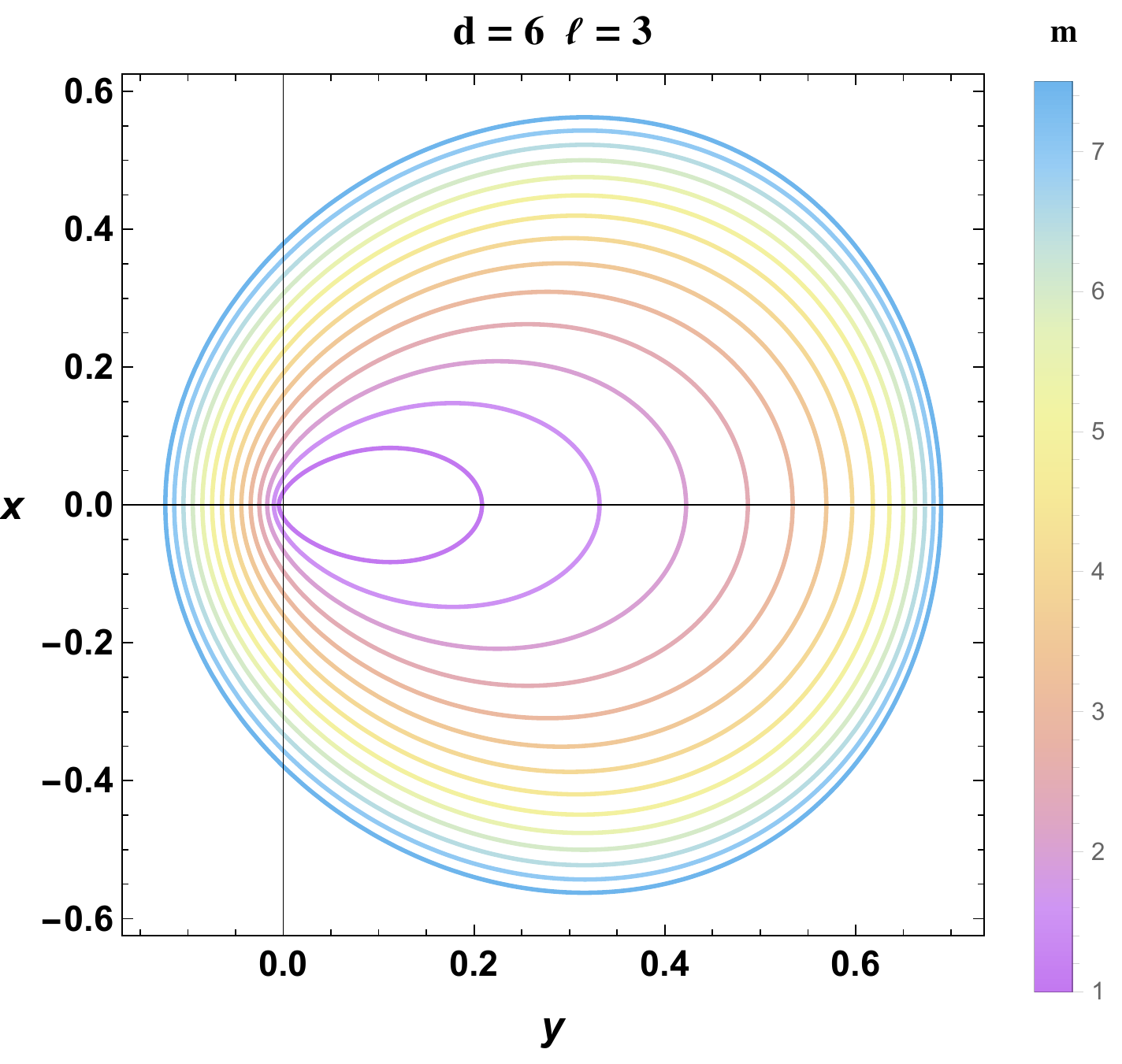}};	
\node[draw, line width=0.5pt,color=red,dashed] at (46.5,1.8){};
\draw[<-,line width=0.6pt,color=red](30.7,2.3)--(45.8,2.3);
\end{tikzpicture}	
\caption{Shadow geometries of superentropic AdS black holes in $d=6$  for different values of $m$ and $\ell$ in four dimensions. The observer is positioned at $r_{ob}=50$ and $\theta_{ob}=\frac{\pi}{2}$.}
\label{ben1}
\end{center}
\end{figure*}

 Here, we study  the shadow behaviors for  $d>5$.   Concretely, we show that the naked singularity  should be evinced.   To reveal that, we need  to solve the equation of $\Delta=0$  in higher dimensions. It has been remarked that it is  complicated to   provide   analytical solutions.  However, we can reveal  that this equation involves at least  a real root which  removes the   naked singularity behaviors. For generic values of $r$, it is obvious that  $\Delta$ is a continuous    radial function. Taking into account the limits   $\lim\limits_{r \to 0^+}\Delta=-\infty$ and  $ \lim\limits_{r \to +\infty}\Delta=+\infty$, we can  safely say that  $\Delta=0$ has a  real solution.  Based on this argument,    the naked singularity  has been removed  for   $d>5$ due to the existence of  the real roots.  To inspect the associated behaviors,   we illustrate the  six dimensional shadow geometries in terms of $m$ and $\ell$. In particular, we  fix one parameter  and vary  the remaining one.     Fig.(\ref{ben1}) provides  the performed computations. 
The left panel illustrates the mass variations, while the right one  gives  $\ell$  variations.
 For  a fixed value of  $\ell$,  we remark that the shadow size   increases with $m$  contrary  to  $\ell$.   Fixing $m$, the shadow size  decreases by increasing  $\ell$ being an expected behavior.\\
\section{Discussion and concluding remarks }
The study of  the optical properties  of the black holes in  higher dimensions   could   bring  global pictures  associated with non-trivial  gravity models including  superstrings and M-theory.
  Motivated   by  such supergravity extended models, we have investigated  the shadow behaviors of  the superentropic black holes in  arbitrary dimensions.  Applying the Hamilton-Jacobi method, we have  first obtained the  null geodesic equations  of motion in terms of the space-time dimension $d$ where   a discussion on the extra direction contributions   has been elaborated.     Considering  a  mass constraint in arbitrary dimensions, we  have generalized the shadow equations   by  exploiting  the higher dimensional  spheric  coordinates. Applying  a  spheric stereographic projection,  we have studied   shadow  behaviors  in terms of  one-dimensional real curves.   The  present  results recover   the previous findings obtained in  four dimensions \cite{14}.          Fixing   the mass parameter, we have investigated   the   shadow  shapes in arbitrary dimensions.  We have  shown that  the  shadow size  decreases  by increasing $d$.   For a fixed value of  $d$,   the  size increases by  decreasing 
 $\ell$.  In addition,  we  have  remarked that $d$ and $\ell$ involve the same effect.  Varying the mass   parameter,  we have  found  nice optical properties. In four dimensions, for instance, we have shown that  the optical behaviors  exhibit transitions  from   the D-shaped elliptic geometry  to  the  naked singularity   via the cardioid curve   associated with  a critical mass value. It has been remarked that the  metric of the elliptic geometry depends on certain parameters. According to  theirs values,  we have discussed  event horizon and horizonless behaviors   \cite{Lamy}.   A close  inspection shows that  a  similar critical  geometry  has been obtained  being called critical curve\cite{cardi1}. Concerning  five dimensions,   this    intermediate    critical geometry  has been evinced, due to the absence of the  mass parameter  in terms of the involved remaining  ones.  In  dimensions more than five,   however, we have  revealed   that the  naked  singularity has  been disappeared.     For higher dimensions,  the  four dimensional  transition   behaviors  have been  removed. \\
 The present work could be compared with certain results associated with  non-rotating and rotating black hole solutions. First, for small values of the $\ell$, the shadow of the superentropic black hole could involve certain similarities  with non-rotating black holes in non-trivial backgrounds. A close examination shows that the obtained shadows could be compared with the ones of  the black holes in the presence of  strong magnetic fields\cite{ca4,ca5}.  Motivated   by these activities,  this could open new windows to     
 provide a  comparative discussion  with Kerr and Kerr-like black holes.  It  has been observed certain distinctions  associated with the shadow size.   Such  distinctions  could be understood   from the involved solutions   where  the mass parameter is constrained with other involved ones. In this  way, it  has been not considered as a free parameter. These constraints  many   provide certain  differences  compared to  known  black hole properties including  thermodynamic and optical aspects. Coming back to the transition of  the superentropic black holes in four dimensions,  the cardioid  geometry  has  appeared  as  a relevant  shadow configuration linked to the critical value of the mass parameter. This mass is considered as a primordial  parameter control the thermodynamic aspect. Indeed, many  works show that the thermodynamic behaviors  of the  superentropic black holes  are different than the  trivial solutions\cite{25,26,27,28,29,30}. This difference  has appeared    also in the optical behaviors. Precisely,  we have  remarked  a  distinction   in  the shadow geometries   compared with  ordinary black hole  solutions\cite{ca3,ca8}.  In this way,    the optical transition in  four dimensional could be considered as a relevant difference. In addition,  this optical transition of  the superentropic black holes  could   be exploited to unveil  the optical behaviors  of  certain black holes in non-trivial solutions. We hope to link  such a  transition  with   future  EHT dada to support the present results. 
 
   This work comes up with certain  open questions.  A possible issue concerns the observational data   supporting the four dimensional  shadow geometric transitions. It should be
   interesting  to  examine the  present   behaviors  by considering external  field contributions  including dark energy and dark matter.   It should be interesting to make contact with  the associated   interesting findings.  Moreover,   the  shadows of higher dimensional multi-center black objects, obtained from supergravity theories,  have been explored in \cite{oq}. It would therefore be of interest to try to make contact with such a work.       We  hope to  address   such open  questions in future  works.  \\           
                                                                            
{\bf Acknowledgement}\\
We would like to thank   N. Askour, H. Belmahi,  H. El Moumni and M. B. Sedra for collaborations and discussions  on the related topics. This work is partially supported by the ICTP through AF.


\begin{thebibliography}{10} 
    \bibitem{1}
K. Akiyama and al.,
\textit{First M87 Event Horizon Telescope Results. IV. Imaging the Central
  Supermassive Black Hole},
 Astrophys. J. {\bf L4} (1) (2019) 875, {\tt arXiv:1906.11241}.

\bibitem{2}
K. Akiyama and al.,
\textit{First M87 Event Horizon Telescope Results. V. Imaging the Central
  Supermassive Black Hole},
 Astrophys. J. {\bf L5} (1) (2019) 875.
\bibitem{3}
K. Akiyama and al.,
\textit{First M87 Event Horizon Telescope Results. VI. Imaging the Central
  Supermassive Black Hole},
 Astrophys. J. {\bf L6} (1) (2019) 875.
 \bibitem{5}
  S. W. Hawking and D. N. Page, \textit{Thermodynamics of black holes in anti-de Sitter space}, Commun. Math. Phys. \textbf{87} (4) (1983) 577.
 \bibitem{6}
D. Kubizňák, R. B. Mann, Mae Teo, \textit{Black hole chemistry: thermodynamics
with Lambda}, Class. Quantum Grav. \textbf{34} (2017) 06300, 	{\tt arXiv:1608.06147}.
\bibitem{7}
S.W. Wei, Y.X. Liu, R.B. Mann
and R. B. Mann, \textit{Novel dual relation and constant in Hawking-Page phase transitions}, Phys.Rev. D \textbf{102}  (08)(2020) 104011, {\tt arXiv:2006.11503}.
\bibitem{8}
A. Belhaj, A. El Balali, W. El Hadri, E. Torrente-Lujan,\textit{ On Universal Constants of AdS Black Holes from Hawking-Page Phase Transition}, Physics Letters B \textbf{811}(2020) 135871, {\tt  arXiv:2010.07837}.
\bibitem{9}
S.~W.~Wei, Y.~X.~Liu and R.~B.~Mann, {\it Novel dual relation and constant in Hawking-Page phase transitions}, Phys. Rev. D \textbf{102} (2020) 104011, 
{\tt arXiv:2006.11503}.
\bibitem{99}

 P.  V. P. Cunha, C. A. R. Herdeiro, E. Radu, H. F. Runarsson, \textit{Shadows of Kerr black holes with scalar hair},    Phys. Rev. Lett.  \textbf{115} (2015) 211102, 	{\tt  arXiv:1509.00021}. 


\bibitem{10}
A. Belhaj, M. Benali, A. El Balali, H. El Moumni,  S-E. Ennadifi,
\textit{Deflection angle and shadow behaviors of quintessential black holes in arbitrary dimensions}, Class. Quant.  Grav. \textbf{37} (2020) 215004, {\tt arXiv:2006.01078}.
\bibitem{11}
Z.~Li and J.~Jia, {\it Kerr-Newman-Jacobi geometry and the deflection of charged massive particles}, Phys. Rev. D \textbf{104} (2021) 04406, {\tt arXiv:2108.05273}.
\bibitem{12}
S.~I.~Kruglov, {\it New model of 4D Einstein--Gauss--Bonnet gravity coupled with nonlinear electrodynamics}, Universe \textbf{7} (2021) 249, {\tt arXiv:2108.07695}.
\bibitem{13}
K.~Matsuno, {\it Light deflection by squashed Kaluza-Klein black holes in a plasma medium}, Phys. Rev. D \textbf{103} (2021) no.4, 044008, {\tt arXiv:2011.07742}.

\bibitem{ca1}
P.~V.~P.~Cunha and C.~A.~R.~Herdeiro, {\it Shadows and strong gravitational lensing: a brief review},
Gen. Rel. Grav. \textbf{50} (2018) 42, {\tt arXiv:1801.00860}.


\bibitem{ca2}
P.~V.~P.~Cunha, C.~A.~R.~Herdeiro, B.~Kleihaus, J.~Kunz and E.~Radu, {\it Shadows of Einstein\textendash{}dilaton\textendash{}Gauss\textendash{}Bonnet black holes},
Phys. Lett. B \textbf{768} (2017) 373, {\tt arXiv:1701.00079}. 

\bibitem{ca6}
M.~Okyay and A.~\"Ovg\"un, {\it Nonlinear electrodynamics effects on the black hole shadow, deflection angle, quasinormal modes and greybody factors}, JCAP \textbf{01} (2022)  009, {\tt arXiv:2108.07766}.

\bibitem{ca7}
C.~A.~R.~Herdeiro, {\it Black holes: on the universality of the Kerr hypothesis}, {\tt arXiv:2204.05640}.


\bibitem{ca9}
H.~C.~D.~Lima, Junior., L.~C.~B.~Crispino, P.~V.~P.~Cunha and C.~A.~R.~Herdeiro, {\it Can different black holes cast the same shadow?}, Phys. Rev. D \textbf{103} (2021) 084040, {\tt arXiv:2102.07034}.

\bibitem{ca10}
C.~A.~R.~Herdeiro, A.~M.~Pombo, E.~Radu, P.~V.~P.~Cunha and N.~Sanchis-Gual, {\it The imitation game: Proca stars that can mimic the Schwarzschild shadow}, JCAP \textbf{04} (2021) 051, {\tt arXiv:2102.01703}. 


\bibitem{ca11}
P.~V.~P.~Cunha and C.~A.~R.~Herdeiro, {\it Stationary black holes and light rings}, Phys. Rev. Lett. \textbf{124} (2020) 181101, {\tt arXiv:2003.06445}.

\bibitem{ca12}
A.~\"Ovg\"un and \.I.~Sakall\i{}, {\it Testing generalized Einstein\textendash{}Cartan\textendash{}Kibble\textendash{}Sciama gravity using weak deflection angle and shadow cast}, Class. Quant. Grav. \textbf{37} (2020) 225003, {\tt arXiv:2005.00982}. 





\bibitem{14}
A.~Belhaj, H.~Belmahi, M.~Benali, {\it Superentropic AdS black hole shadows}, Phys. Lett. B \textbf{821} (2021) 136619, {\tt arXiv:2110.06771}.
\bibitem{15}
A. Grenzebach, V. Perlick, C. L\"{a}mmerzahl, \textit{Photon Regions and Shadows of Accelerated Black Holes},  Int. J. Mod. Phys. D{\bf 24} (09) (2015) 1542024, {\tt arXiv:1503.03036}.
\bibitem{16}
A. Grenzebach, V. Perlick,  C. L\"{a}mmerzahl, \textit{Photon Regions and Shadows of Kerr-Newman-NUT Black Holes with a Cosmological Constant}, Phys. Rev. D {\bf 89} (12) (2014) 124004, {\tt arXiv:1403.5234}.
\bibitem{17}
A. de Vries,
 \textit{The apparent shape of a rotating charged black hole, closed photon orbits and the bifurcation set A4}, Class. Quant. Grav. 17, (2000)123.
  \bibitem{18}
A.~Övgün, I.~Sakalli, J.~Saavedra,
\textit{Shadow cast and Deflection angle of Kerr-Newman-Kasuya spacetime}, JCAP \textbf{10} (2018) 041, {\tt arXiv:1807.00388}.
  \bibitem{20}
 A. Belhaj, H. Belmahi,  M. Benali, W. El Hadri, H. El Moumni,  E. Torrente-Lujan,  \textit{Shadows of 5D Black Holes from string theory},  Phys. Lett.
B \textbf{812} (2021) 136025, {\tt arXiv:2008.13478}.
 \bibitem{21}
J. R. Farah, D. W. Pesce, M. D. Johnson, L. Blackburn, {\it On the Approximation of the Black Hole Shadow with a Simple Polar Curve}, ApJ {\bf 900}  (2020) 77, {\tt arXiv:2007.06732}.
  \bibitem{200} A. Belhaj, M. Benali, A. El Balali, W. El Hadri, H. El Moumni, E. Torrente-Lujan,  \textit{Black hole shadows in M-theory scenarios}, Int. J. Mod. Phys. D  \textbf{30} (2021) 2150026,
{\tt arXiv:2008.09908}.
 \bibitem{22}
G. C. Bower, A. Deller, P. Demorest, A. Brunthaler, H. Falcke, M. Moscibrodzka, R. P. Eatough, M. Kramer, K. J. Lee,  L. Spitler, et al. {\it The Proper Motion of
the Galactic Center Pulsar Relative to Sagittarius A*}, Astrophys. J.  \textbf{798} (2015) 2, {\tt arXiv:1411.0399}.
 \bibitem{23}
P. Kocherlakota, S. Biswas, P. S. Joshi, S. Bhattacharyya, C. Chakraborty,  A. Ray, {\it Gravitomagnetism and Pulsar Beam Precession near a Kerr Black Hole}, Mon. Not. Roy. Astron. Soc. {\bf 490} (2019) 3, {\tt arXiv:1711.04053}.
\bibitem{25}
R. A. Hennigar, D. Kubiznak,  R. B. Mann, \textit{Entropy
Inequality Violations from Ultraspinning Black Holes}, Phys. Rev. Lett. \textbf{115}(2015)  031101,
{\tt arXiv:1411.4309}.
 \bibitem{230}
R.  A. Hennigar, D. Kubiznak,  R. B. Mann,   \textit{Super-Entropic Black Holes},  Phys. Rev. Lett. {\bf 115}(2015) 031101,  {\tt  arXiv:1411.4309}.
\bibitem{24}
D. Klemm, \textit{Four-dimensional black holes with unusual
horizons}, Phys.Rev. D \textbf{89} (2014) 084007, {\tt
arXiv:1401.3107}.

\bibitem{26}
R A. Hennigar, D Kubiznak, R. B. Mann, N. Musoke, \textit{Ultraspinning limits and super-entropic black holes}, JHEP \textbf{096} (2015) 1506, {\tt arXiv:1504.07529}.

\bibitem{27}
R.A. Hennigar, D. Kubiznak,  R.B. Mann, \textit{Super-entropic black holes}, Phys. Rev. Lett. \textbf{115} (2015) 031101, {\tt arXiv:1411.4309}.

\bibitem{28}
D. Wu, P. Wu, H. Yu, S. Q. Wu \textit{Are ultra-spinning Kerr-Sen-AdS4 black holes always super-entropic ?},
Phys. Rev. D \textbf{102}  (2020) 044007, {\tt arXiv:2007.02224}.

\bibitem{29}
D. Wu, P. Wu, H. Yu, S. Q. Wu, \textit{Notes on thermodynamics of super-entropic AdS black holes}, Phys. Rev.D  \textbf{101}  (2020) 024057, {\tt arXiv:1912.03576}.
\bibitem{30}
D. Wu, S.Q. Wu, P. Wu, H. Yu,
\textit{Aspects of the dyonic Kerr-Sen-AdS black hole and its ultraspinning version}, Phys. Rev. D \textbf{103} (2021) 044014, {\tt arXiv:2010.13518}.
\bibitem{31}
S. Chandrasekhar,
 \textit{The mathematical theory of black holes}, volume~69.
 Oxford University Press, 1998.
\bibitem{31aa}
 K. Jusufi, M. Jamil, P. Salucci, T. Zhu, S. Haroon, {\it Black hole surrounded by a dark matter halo in the m87 galactic center and its identification with shadow images}, Phys. Rev. D {\bf100}  (2019) 044012, {\tt arXiv:1905.11803}.
 \bibitem{32}
   L. Amarilla, E. F. Eiroa,
  \textit{Shadow of a rotating braneworld black hole}, Phys. Rev. D {\bf85} (2012) 064019, {\tt arXiv:1112.6349}.

\bibitem{Lamy}
F.~Lamy, E.~Gourgoulhon, T.~Paumard, F.~H.~Vincent, {\it Imaging a non-singular rotating black hole at the center of the Galaxy},
Class. Quant. Grav. \textbf{35} (2018) 115009, {\tt arXiv:1802.01635}.
 \bibitem{cardi1}
O.~James, E.~von Tunzelmann, P.~Franklin,  K.~S.~Thorne, {\it Gravitational Lensing by Spinning Black Holes in Astrophysics, and in the Movie Interstellar},
Class. Quant. Grav. \textbf{32} (2015)6, {\tt arXiv:1502.03808}.  

\bibitem{ca4}
H.~C.~D.~L.~Junior, P.~V.~P.~Cunha, C.~A.~R.~Herdeiro and L.~C.~B.~Crispino, {\it Shadows and lensing of black holes immersed in strong magnetic fields},
Phys. Rev. D \textbf{104} (2021)  044018, {\tt arXiv:2104.09577}. 


\bibitem{ca5}
H.~C.~D.~L.~Junior, J.~Z.~Yang, L.~C.~B.~Crispino, P.~Cunha, V.P. and C.~A.~R.~Herdeiro,
{\it Einstein-Maxwell-dilaton neutral black holes in strong magnetic fields: Topological charge, shadows, and lensing}, Phys. Rev. D \textbf{105} (2022)  064070, {\tt arXiv:2112.10802}. 
\bibitem{ca3}
P.~Cunha, V.P., C.~A.~R.~Herdeiro and E.~Radu, {\it Spontaneously Scalarized Kerr Black Holes in Extended Scalar-Tensor\textendash{}Gauss-Bonnet Gravity}, Phys. Rev. Lett. \textbf{123} (2019) 011101, {\tt arXiv:1904.09997}.
\bibitem{ca8}
R.~C.~Pantig, P.~K.~Yu, E.~T.~Rodulfo and A.~\"Ovg\"un, {\it Shadow and weak deflection angle of extended uncertainty principle black hole surrounded with dark matter}, {\tt arXiv:2104.04304}.
\bibitem{oq}T. Hertog, T. Lemmens, B. Vercnocke, {\it  Imaging Higher Dimensional Black Objects}, Phys. Rev. D  \textbf{100}  (2019) 046011, {\tt arXiv:1903.05125}.

 
 \end{thebibliography}
\end{document}